\documentclass[11pt,a4paper]{article}
\usepackage{amsmath,amssymb}
\usepackage{epsfig,graphicx}
\begin{document}
\title{\bf Light Scalars in Field Theory \thanks{Invited talk at
 QUARKS'2008, Sergiev Posad, Russia, May 23-29, 2008}}
\author{N.N. Achasov
\footnote{{\bf e-mail address}: achasov@math.nsc.ru}\\
 \small{\em
 Sobolev Institute for Mathematics}\\
\small{\em Academician Koptiug Prospekt, 4,
  Novosibirsk, 630090, Russia}}
  \date{}
 \maketitle

\begin{abstract}
Outline: 1. Introduction, 2. Confinement, chiral dynamics and
light scalar mesons, 3. Chiral shielding of  $\sigma(600)$, chiral
constraints,  $\sigma(600)$, $f_0(980)$ and their mixing in
$\pi\pi\to\pi\pi$, $\pi\pi\to K\bar K$,  and
$\phi\to\gamma\pi^0\pi^0$, 4. The $\phi$ meson radiative decays on
light scalar resonances. 5. Why   $a_0(980)$ and  $f_0(980)$ are
not the $K\bar K$ molecules. 6. Light scalars in $\gamma\gamma$
collisions.

Evidence for four-quark components of light scalars is given. The
priority of Quantum Field Theory in revealing the light scalar
mystery is emphasized.
\end{abstract}

\section{Introduction}
The scalar channels in the region up to 1 GeV became a stumbling
block of QCD. The point is that both perturbation theory and sum
rules do not work in these channels because there are not solitary
resonances in this region.

At the same time the question on the nature of the light scalar
mesons is major for  understanding the mechanism of the chiral
symmetry realization, arising from the confinement, and hence for
understanding the confinement itself.

\section{Place in QCD}
The QCD Lagrangian is given by \\ $L$=$-\frac{1}{2}Tr\left
(G_{\mu\nu}(x)G^{\mu\nu}(x)\right)+\bar q(x)(i\hat{D}-M)q(x),$\\
$M$ is a diagonal matrix of quark masses, $\hat{D}$=$\gamma^\mu
D_\mu, D_\mu$=$\partial_\mu+ig_0G_\mu (x).$ $M$ mixes left and
right spaces. But in chiral limit, $M_{ff}$\,$\to$\,0, these
spaces separate realize $U_L(3)\times U_R(3)$ flavour symmetry,
which, however, is broken by the gluonic anomaly up to
$U_{\mbox{\scriptsize{vec}}}(1)\times SU_L(3)\times SU_R(3).$ As
experiment suggests, confinement forms colourless observable
hadronic fields and spontaneous breaking of chiral symmetry with
massless pseudoscalar fields. There are two possible scenarios for
QCD at low energy. 1. Non-linear $\sigma$ model. 2. Linear
$\sigma$ model (LSM). The experimental nonet of the light scalar
mesons, $f_0(600)$ (or $\sigma (600)$), $\kappa$(700-900),
$a_0(980)$ and $f_0(980)$ mesons, suggests the $U_L(3)\times
U_R(3)$ LSM.
\section{History}
Hunting the light  $\sigma$ and $\kappa$ mesons had begun in the
sixties already and a preliminary information on the light scalar
mesons in 
PDG Reviews had appeared at that time. But long-standing
unsuccessful attempts to prove their existence in a conclusive way
entailed general disappointment and an information on these states
disappeared from PDG Reviews. One of principal reasons against the
$\sigma$ and $\kappa$ mesons was the fact that both $\pi\pi$ and
$\pi K$ scattering phase shifts do not pass over $90^0$ at
putative resonance masses.
\section{\boldmath $SU_L(2)\times SU_R(2)$ LSM,
$\pi\pi\to\pi\pi$ \cite{GML,AS94,AS07}} Situation changes when we
showed that in LSM there is a negative background phase which
hides the $\sigma$ meson in $\pi\pi\to\pi\pi$. It has been made
clear that shielding wide lightest scalar mesons in chiral
dynamics is very natural. This idea was picked up and triggered
new wave of theoretical and experimental searches for the $\sigma$
and $\kappa$ mesons. Our approximation is as follows (see Fig.1):
\begin{figure}[h]
\includegraphics[width=14cm]{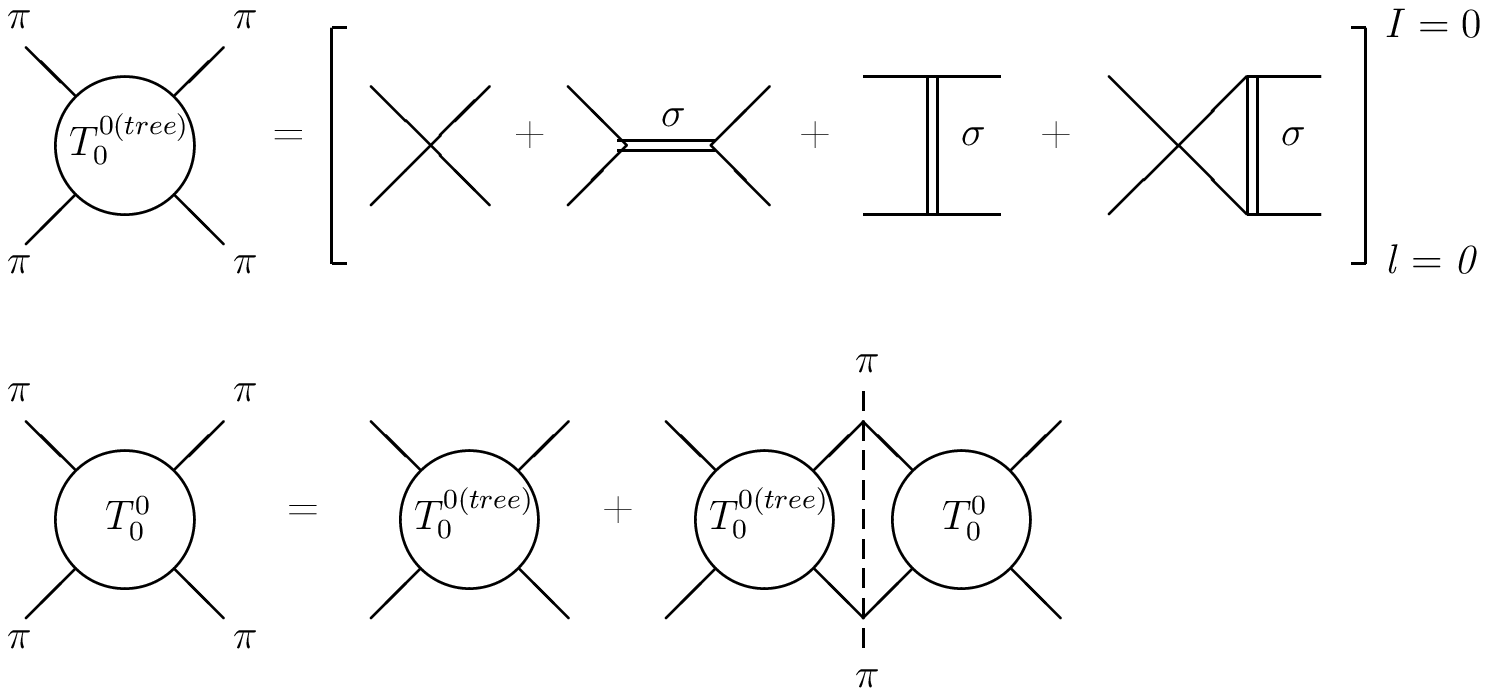}\\
{\footnotesize Figure 1. The graphical representation of the $S$
wave $I=0$ $\pi\pi$ scattering amplitude $T^0_0$.}
\end{figure}

$T^0_0=\frac{T_0^{0(tree)}}{1-i\rho_{\pi\pi}T_0^{0(tree)}}$=
$\frac{e^{2i\delta^0_0}-1}{2i\rho_{\pi\pi}}$=
$\frac{e^{2i(\delta_{bg}+\delta_{res})}-1}{2i\rho_{\pi\pi}},\\[9pt]
T^2_0=\frac{T_0^{2(tree)}}{1-i\rho_{\pi\pi}T_0^{2(tree)}}=
\frac{e^{2i\delta_0^2}-1}{2i\rho_{\pi\pi}}$.

\section{Results in our approximation \cite{AS07}.}
$M_{res}=0.43$ GeV, $\Gamma_{res}(M^2_{res})=0.67$ GeV, $
m_\sigma =0.93$ GeV, \\[9pt]
$\Gamma_{res}(s)=\frac{g_{res}^2(s)}{16\pi\sqrt{s}}\rho_{\pi\pi}\,,\
g_{res}(M^2_{res})/g_{\sigma\pi\pi}=0.33$, \\[9pt]
$a^0_0=0.18\, m_\pi^{-1}\,,\ a^2_0$=$-0.04\, m_\pi^{-1}\,,\
(s_A)^0_0$=$0.45\, m^2_\pi\,,\ (s_A)^2_0$=$2.02\, m^2_\pi$.
\section{Chiral shielding in \boldmath $\pi\pi\to\pi\pi$ \cite{AS07}}
\begin{figure}[h]
\includegraphics[width=14cm]{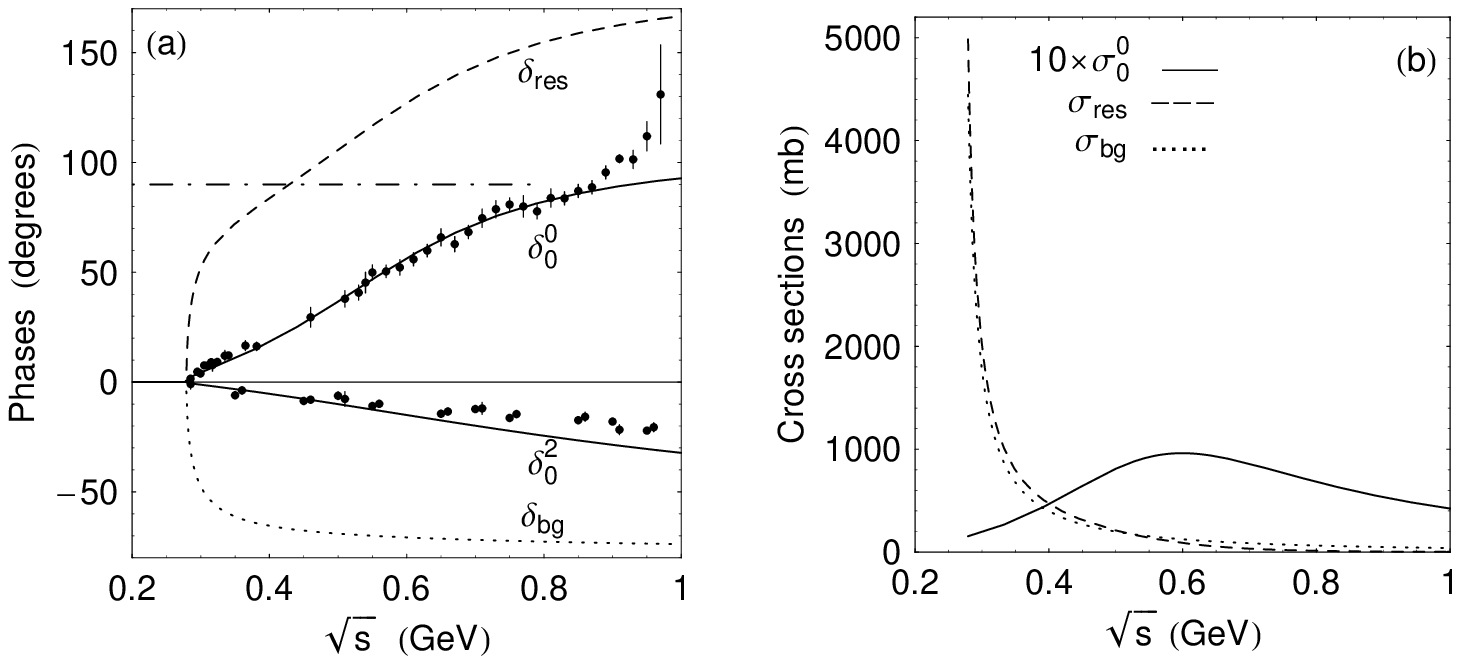}\\
 {\footnotesize Figure 2. The $\sigma$
model. Our approximation. $\delta^0_0=\delta_{res}+\delta_{bg}$.
($\sigma^0_0,\sigma_{res},\sigma_{bg}$)=$32\pi(|T^0_0|^2,
|T_{res}|^2,|T_{bg}|^2)/s$.}
\end{figure}
The chiral shielding of the $\sigma(600)$ meson is illustrated in
Fig. 2.
\section{The \boldmath $\sigma$ pole in $\pi\pi\to\pi\pi$ \cite{AS07}}
$T^0_0\to g^2_\pi/(s-s_R)\,,\  g^2_\pi=(0.12+i0.21)$
GeV$^2$,\\[9pt]
$\sqrt{s_R}=M_R-i\Gamma_R/2=(0.52-i0.25)$ GeV.\\[9pt]
 Considering the
residue of the $\sigma$ pole in $T^0_0$ as the square of its
coupling constant to the $\pi\pi$ channel is not a clear guide to
understand the $\sigma$ meson nature for its great obscure
imaginary part.

\section{The \boldmath{$\sigma$} propagator \cite{AS07}}
$1/D_\sigma(s) = 1/[M^2_{res}- s + \mbox{Re}\Pi_{res}(M^2_{res})-
\Pi_{res}(s)]$.\\[9pt]
  The $\sigma$ meson self-energy $\Pi_{res}(s)$
is caused by the intermediate $\pi\pi$ states, that is, by the
four-quark intermediate states if we keep in mind that the
$SU_L(2)\times SU_R(2)$ LSM could be the low energy realization of
the two-flavour QCD. This contribution shifts the Breit-Wigner
(BW) mass greatly $m_\sigma - M_{res}=0.50$ GeV. So, half the BW
mass is determined by the four-quark contribution at least. The
imaginary part dominates the propagator modulus in the region
300\,MeV\,$<\sqrt{s}<$\,600\, MeV. So, the $\sigma$ field is
described by  its four-quark component at least in this energy
region.
\section{Chiral shielding in \boldmath $\gamma\gamma\to\pi\pi$ \cite{AS07}}
$T_S(\gamma\gamma\to\pi^+\pi^-)=T_S^{Born}(\gamma\gamma\to\pi^+\pi^-)+8\alpha
I_{\pi^+\pi^-}T_S(\pi^+\pi^-\to\pi^+\pi^-)\,, \\[9pt]
T_S(\gamma\gamma\to\pi^0\pi^0)=8\alpha
I_{\pi^+\pi^-}\,T_S(\pi^+\pi^-\to\pi^0\pi^0)\,,\\[9pt]
 T_S^{Born}(\gamma\gamma\to\pi^+\pi^-)=
(8\alpha/\rho_{\pi^+\pi^-})\mbox{Im}I_{\pi^+\pi^-}\,\
I_{\pi^+\pi^-}= \frac{m^2_\pi}{s}
(\pi+i\ln\frac{1+\rho_{\pi\pi}}{1-\rho_{\pi\pi}})^2-1,\ \ s\geq
4m_\pi^2\,.$\\[9pt]
Our results are shown in Fig. 3.

\begin{figure}[h]
\includegraphics[width=14cm]{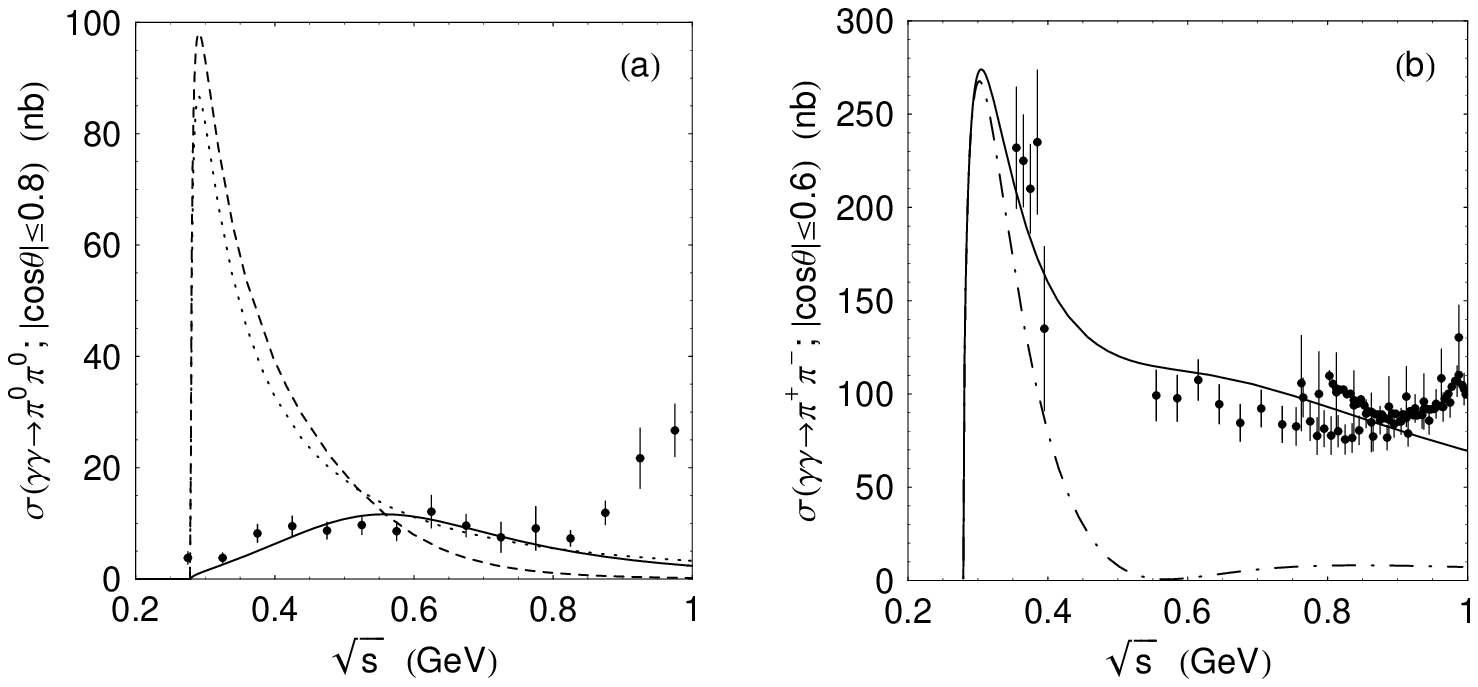} \\
{\footnotesize Figure 3. (a) The solid, dashed, and dotted lines
are $\sigma_S(\gamma\gamma\to\pi^0\pi^0)$,
$\sigma_{res}(\gamma\gamma\to\pi^0\pi^0)$, and
$\sigma_{bg}(\gamma\gamma\to\pi^0\pi^0)$. (b) The dashed-dotted
line is $\sigma_S(\gamma\gamma\to\pi^+\pi^-)$, the solid line
includes the higher waves from
$T^{Born}(\gamma\gamma\to\pi^+\pi^-)$.}
\end{figure}
\vspace*{8pt} $\Gamma(\sigma\to\pi^+\pi^-\to\gamma\gamma\,,s)=
\frac{1}{16\pi\sqrt{s}}|g(\sigma\to\pi^+\pi^-\to\gamma\gamma,s)|^2$,\\[9pt]
 where
$g(\sigma$\,$\to$\,$\pi^+\pi^-$\,$\to$\,$\gamma\gamma,\,s)$=$
(\alpha/2\pi)$ $\times I_{\pi^+\pi^-}g_{res\,\pi^+\pi^-}(s)$; see
Fig. 4. So, the the $\sigma\to\gamma\gamma$ decay is described by
the triangle $\pi^+\pi^-$ loop diagram
$res\to\pi^+\pi^-\to\gamma\gamma$. Consequently, it is due to the
four-quark transition because we imply a low energy realization of
the two-flavour QCD by means of the the $SU_L(2)\times SU_R(2)$
LSM.
\begin{figure}[h]
\includegraphics[width=10cm,height=5cm]{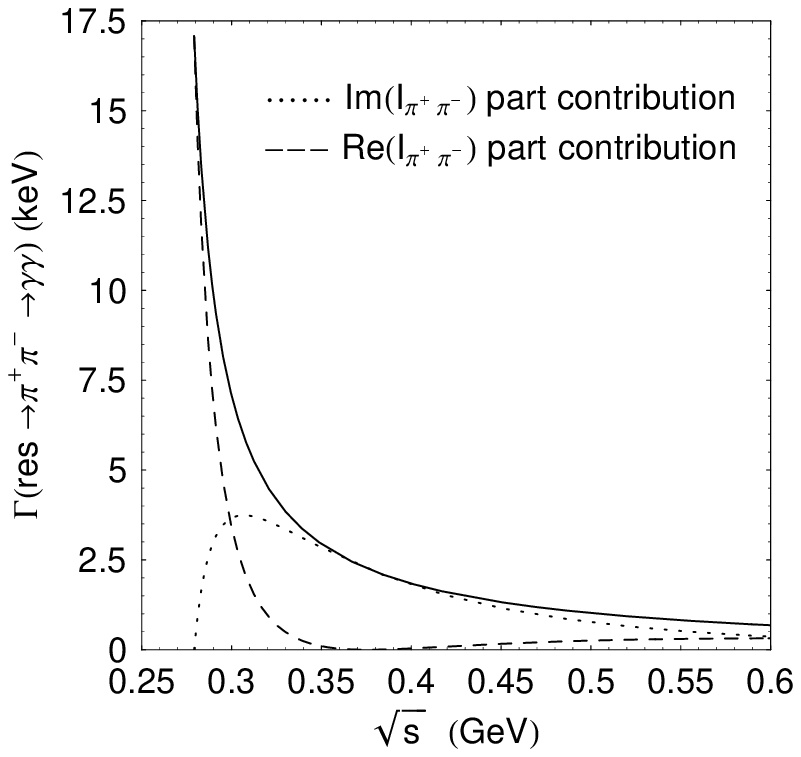}\\
{\footnotesize Figure 4. The energy dependent width of the
$\sigma\to\pi^+\pi^- \to\gamma\gamma$ decay.}\end{figure}

 As Fig. 4 suggests, the real intermediate $\pi^+\pi^-$ state
dominates in $g(res\to\pi^+\pi^-\to\gamma\gamma)$ in the $\sigma$
region $\sqrt{s}<$\,0.6\,GeV. Thus the picture in the physical
region is clear and informative. But, what  about the pole in the
complex $s$ plane? Does the pole residue reveal the $\sigma$
indeed?
\section{The \boldmath $\sigma$  pole in $\gamma\gamma\to\pi\pi$ \cite{AS07}}
$\frac{1}{16\pi}\sqrt{\frac{3}{2}}\,T_S(\gamma\gamma\to\pi^0\pi^0)\to
g_\gamma g_\pi/(s-s_R)$,\\[9pt]
$g_\gamma g_\pi=(-0.45 - i0.19)\cdot 10^{-3}\,\mbox{GeV}^2$,
$g_\gamma/g_\pi= (-1.61+i1.21)\cdot10^{-3}$,
$\Gamma(\sigma\to\gamma\gamma)$=$\frac{|g_\gamma |^2}{M_R}\approx
2\,\mbox{keV}.$\\[9pt]
 It is hard to believe that anybody could learn
the complex but physically clear dynamics of the
$\sigma\to\gamma\gamma$ decay from the residues of the $\sigma$
pole.
\section{Discussion \cite{AS07,CCL,Ja,ADS}}
Leutwyler and collaborators obtained $\sqrt{s_R}= M_R-i\Gamma_R/2
=\left(441^{+16}_{-8}-i272^{+12.5}_{-9}\right )\,\mbox{MeV}$. Our
result agrees with the above only qualitatively, $\sqrt{s_R}=
M_R-i\Gamma_R/2 =(518-i250)\,\mbox{MeV}$. It is natural, for our
approximation.

Could the above scenario incorporates the primary lightest scalar
Jaffe four-quark  state? Certainly the direct coupling of this
state to $\gamma\gamma$ via neutral vector pairs ($\rho^0\rho^0$
and $\omega\omega$), contained in its wave function, is
negligible, $\Gamma(q^2\bar
q^2\to\rho^0\rho^0+\omega\omega\to\gamma\gamma) \approx 10^{-3}$
keV, as we showed in 1982 \cite{ADS}. But its coupling to $\pi\pi$
is strong and leads to $\Gamma(q^2\bar
q^2\to\pi^+\pi^-\to\gamma\gamma)$ similar to
$\Gamma(res\to\pi^+\pi^-\to\gamma\gamma)$ in the above Fig. 4.

Let us add to $T_S(\gamma\gamma\to\pi^0\pi^0)$ the amplitude for
the the direct coupling of $\sigma$ to $\gamma\gamma$ conserving
unitarity $T_{direct}(\gamma\gamma\to\pi^0\pi^0)=sg^{(0)}_{\sigma
\gamma\gamma}g_{res}(s)e^{i\delta_{bg}}/D_{res}(s)$. Fitting the
$\gamma\gamma\to\pi^0\pi^0$ data gives a negligible value of
$g^{(0)}_{\sigma\gamma\gamma}$,
$\Gamma^{(0)}_{\sigma\gamma\gamma}=
|M^2_{res}g^{(0)}_{\sigma\gamma\gamma}|^2/(16\pi M_{res})\approx
0.0034$ keV, in astonishing agreement with our prediction
\cite{ADS}.
\section{Phenomenological chiral shielding \cite{AK06}}
$g_{\sigma\pi^+\pi^-}^2/4\pi$=0.99 GeV$^2$,$\ g_{\sigma
K^+K^-}^2/4\pi$=2$\cdot10^{-4}$ GeV$^2$,
$g_{f_0\pi^+\pi^-}^2/4\pi$=0.12 GeV$^2$, $g_{f_0
K^+K^-}^2/4\pi$=1.04 GeV$^2$. The BW masses and width:
$m_{f_0}$=989 MeV, $m_\sigma$=679 MeV, $\Gamma_\sigma$=498 MeV.
The $l$=$I$=0 $\pi\pi$ scattering length $a^0_0$ =$0.223\,
m^{-1}_{\pi^+}$. Figure 5 illustrates the excellent agreement our
phenomenological treatment with the experimental and theoretical
data.
\begin{figure}[h]
\begin{center}
\begin{tabular}{cc}
\includegraphics[width=14pc]{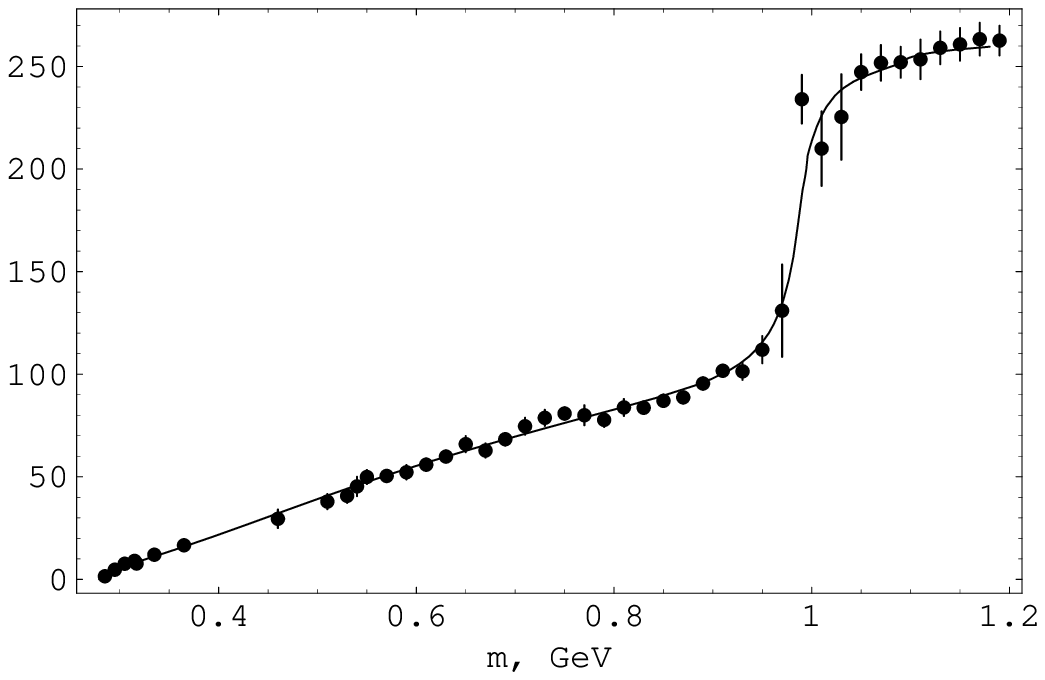}\ \
\includegraphics[width=14pc]{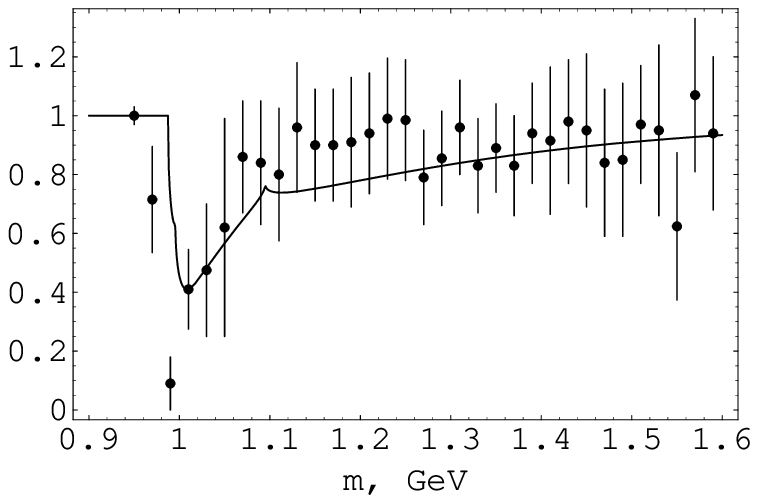}\\
\includegraphics[width=14pc]{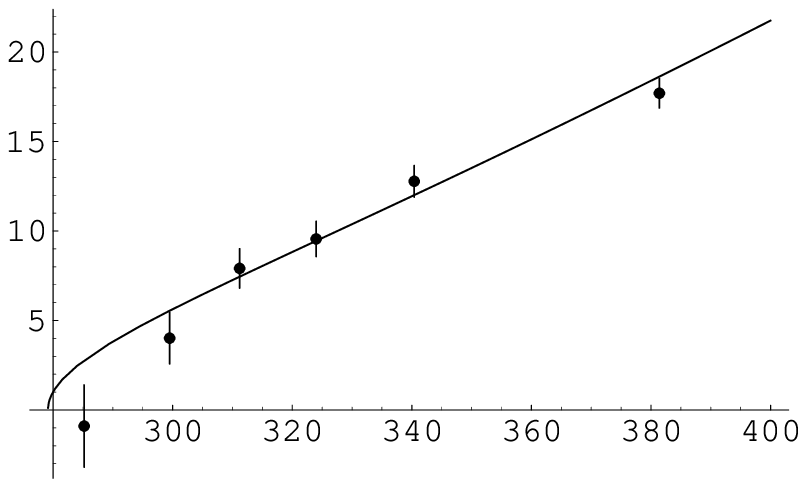}\ \
\includegraphics[width=14pc]{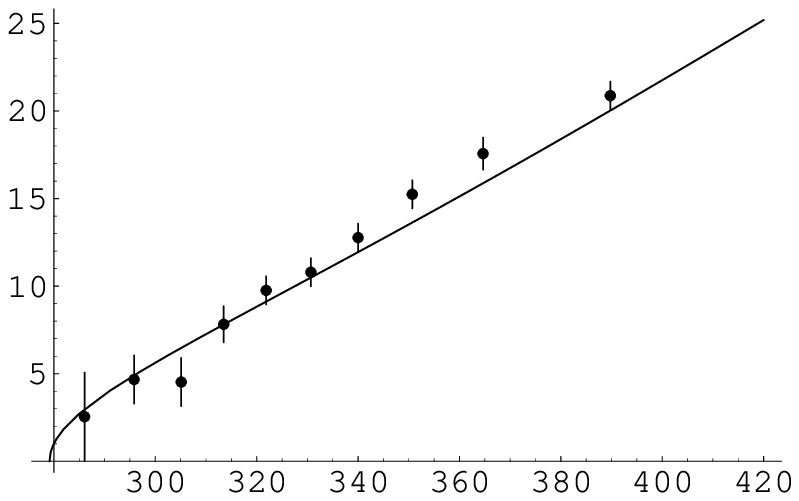}\\
\includegraphics[width=14pc]{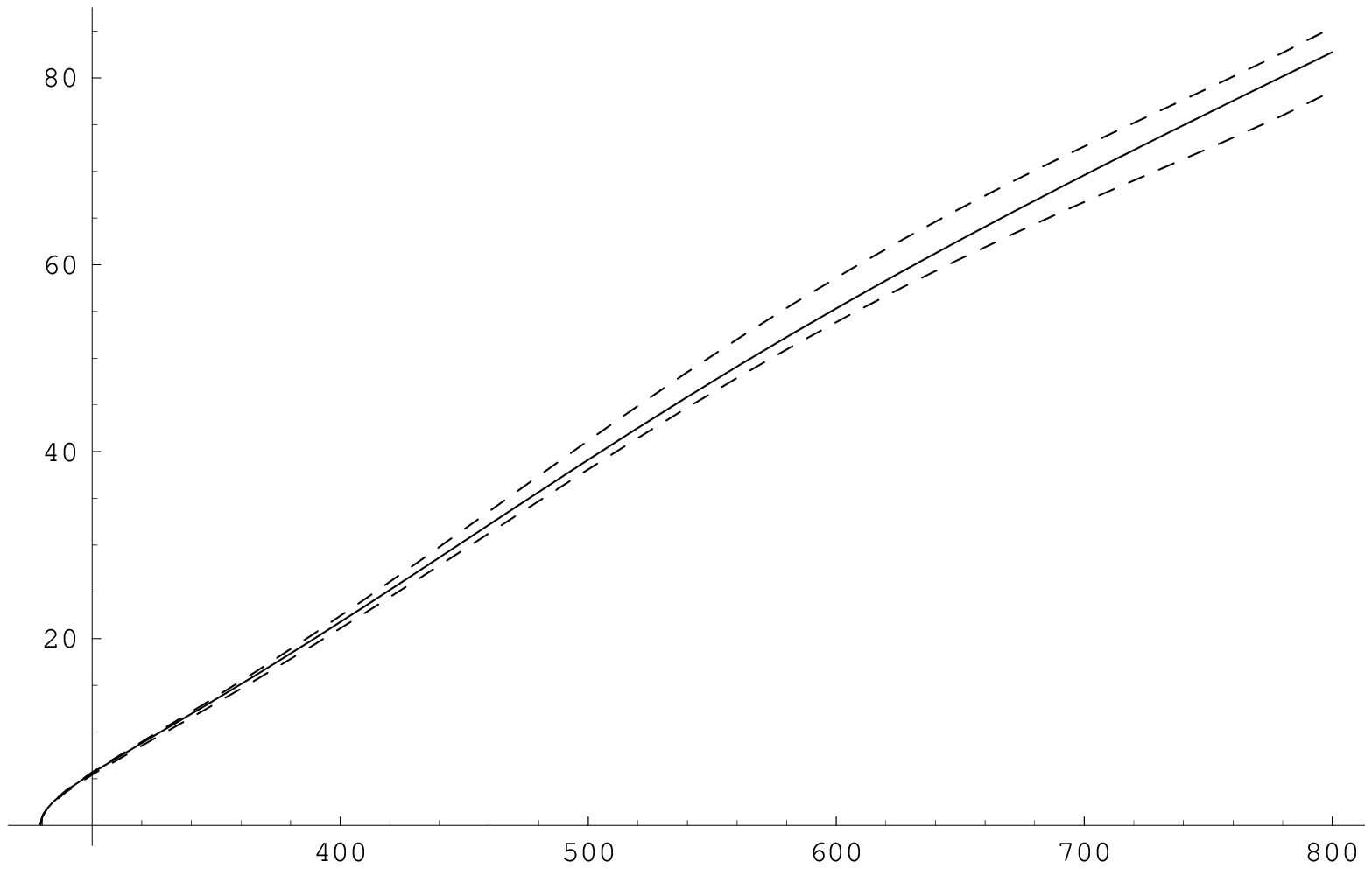}\ \
\includegraphics[width=14pc]{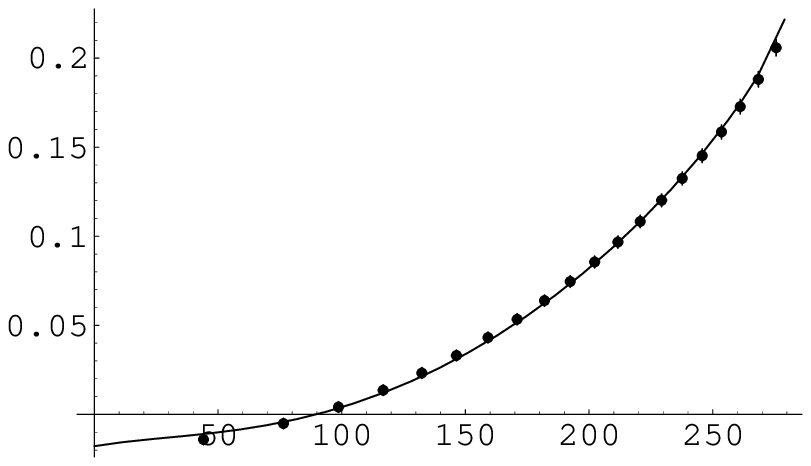}
\end{tabular}\\
\end{center}
{\footnotesize Figure 5. The phenomenological chiral shielding.
$\delta_0^0$=$\delta_B^{\pi\pi}$+$\delta_{res}$. The comparison
with the CERN-Munich data for $\delta^0_0$ and inelasticity
$\eta^0_0$, with the BNL and NA48 data for $\delta^0_0$, with the
CGL band for $\delta^0_0$ \cite{CGL}, and with Leutwyler's
calculation of $T^0_0$ for $s<4m^2_\pi$, respectively.}
\end{figure}
\section{Four-quark model \cite{Ja,A1,A2}}
 There are
numerous evidences in favour of the $q^2\bar q^2$ structure of
$f_0(980)$ and $a_0(980)$. As for the nonet as a whole, even a
dope's look at PDG Review gives an idea of the four-quark
structure of the light scalar meson nonet, $\sigma(600)$,
$\kappa$(700-900), $a_0(980)$, and $f_0(980)$, inverted in
comparison with the classical $P$ wave $q\bar q$ tensor meson
nonet $f_2(1270)$, $a_2(1320)$, $K_2^\ast(1420)$, and $f_2^\prime
(1525)$. Really,  it can be easy understood for the $q^2\bar q^2$
nonet, where $\sigma(600)$ has no strange quarks,
$\kappa$(700-900) has the $s$ quark, $a_0(980)$ and $f_0(980)$
have the $s\bar s$ pair.

\section{Radiative decays of \boldmath $\phi$ meson \cite{AK06,A1,A2,AI,AG9701,AK03}}
Twenty years ago we showed \cite{AI} that the study of the
radiative decays $\phi\to\gamma a_0\to\gamma\pi\eta$ and
$\phi\to\gamma f_0\to \gamma\pi\pi$ can shed light on the problem
of $a_0(980)$ and $f_0(980)$ mesons.  Now these decays have been
studied not only theoretically but also experimentally.  Note that
$a_0(980)$ is produced in the radiative $\phi$ meson decay as
intensively as $\eta '(958)$ containing $\approx 66\% $ of $s\bar
s$, responsible for $\phi\approx s\bar s\to\gamma s\bar s\to\gamma
\eta '(958)$. It is a clear qualitative argument for the presence
of the $s\bar s$ pair in the isovector $a_0(980)$ state, i.e., for
its four-quark nature.
\section{\boldmath
$K^+K^-$ loop mechanism  \cite{AK06,A1,A2,AI,AG9701,AK03}}
 When
basing the experimental investigations, we suggested \cite{AI}
one-loop model $\phi\to K^+K^-\to\gamma\, [a_0(980)/f_0(980)]$;
see Fig. 6.
\begin{figure}[h]
\begin{center}
\begin{tabular}{ccc}
\includegraphics[width=3.5cm]{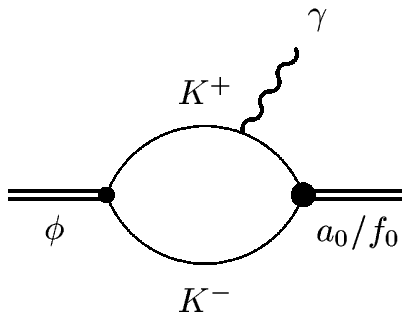}& \raisebox{-6mm}{$\includegraphics[width=3.5cm]{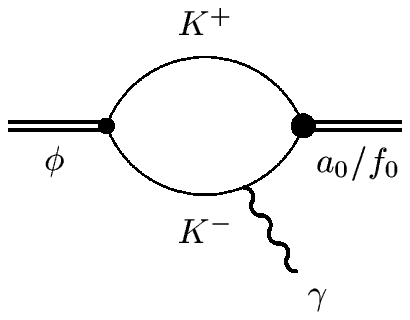}$}&
\includegraphics[width=3.5cm]{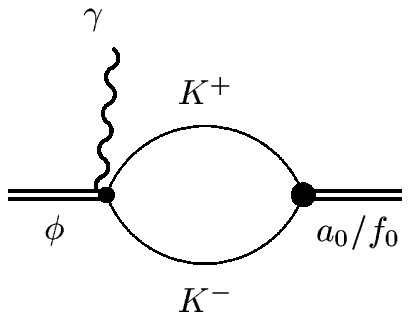}\\ (a)&(b)&(c)
\end{tabular}
\end{center}
{\footnotesize Figure 6. The $K^+K^-$ loop model.}
\end{figure}

This model is used in the data treatment and is ratified by
experiment, see Figs. 7 (a) and 7 (b).
\begin{figure}[h]
\begin{center}
\begin{tabular}{ccc}
\hspace*{-14pt}\includegraphics[width=13pc]{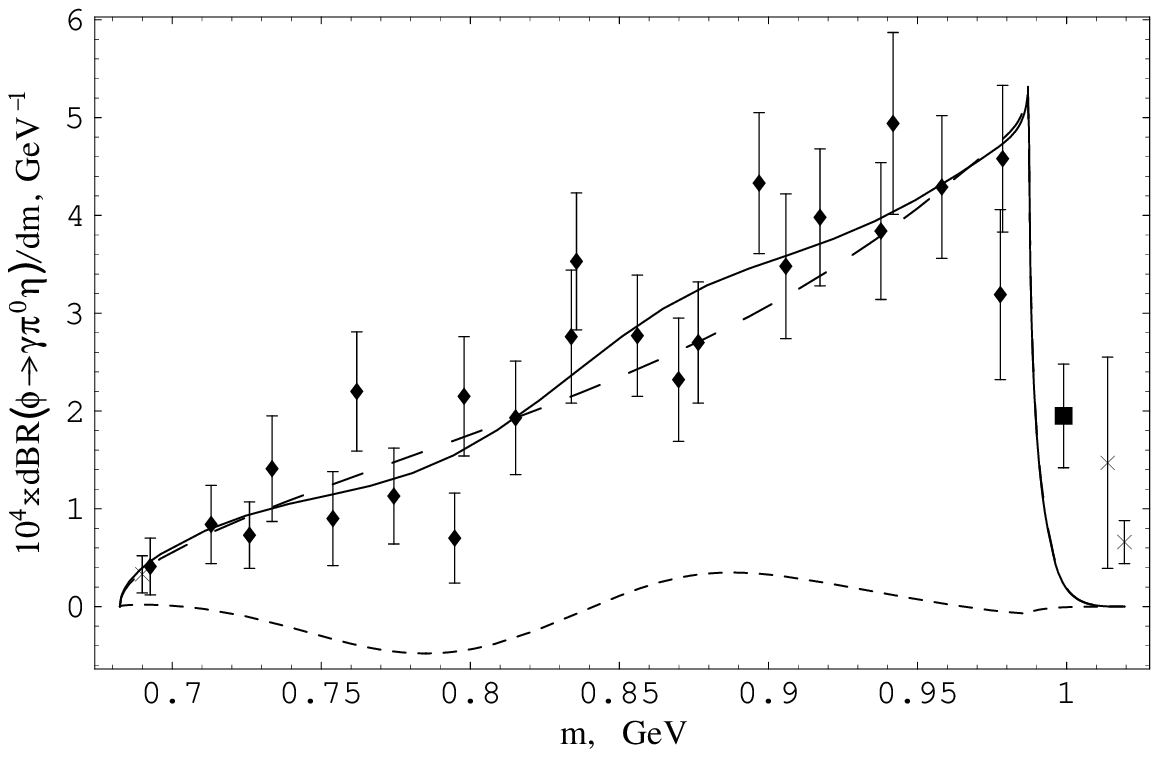}&
\hspace*{-12pt}\includegraphics[width=13pc]{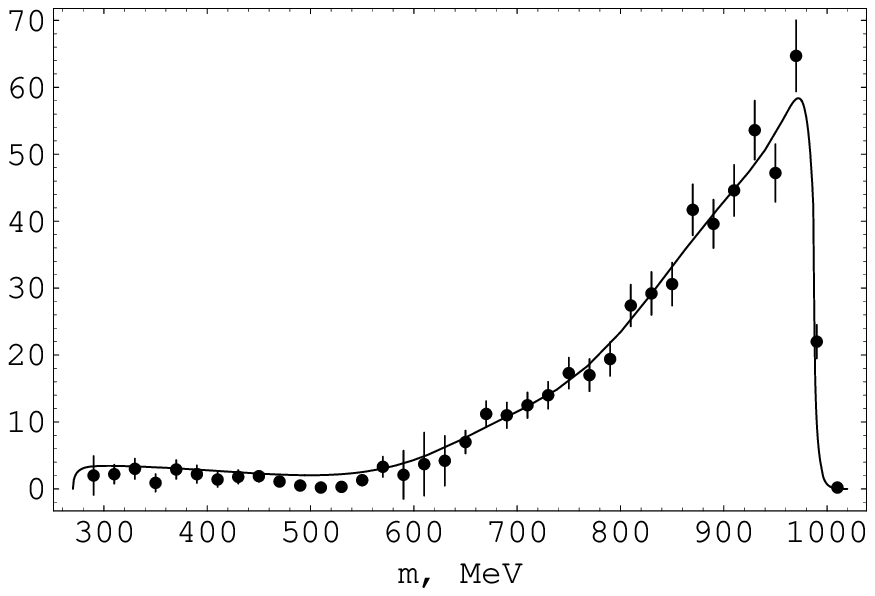}&
\hspace*{-12pt}\includegraphics[width=13pc]{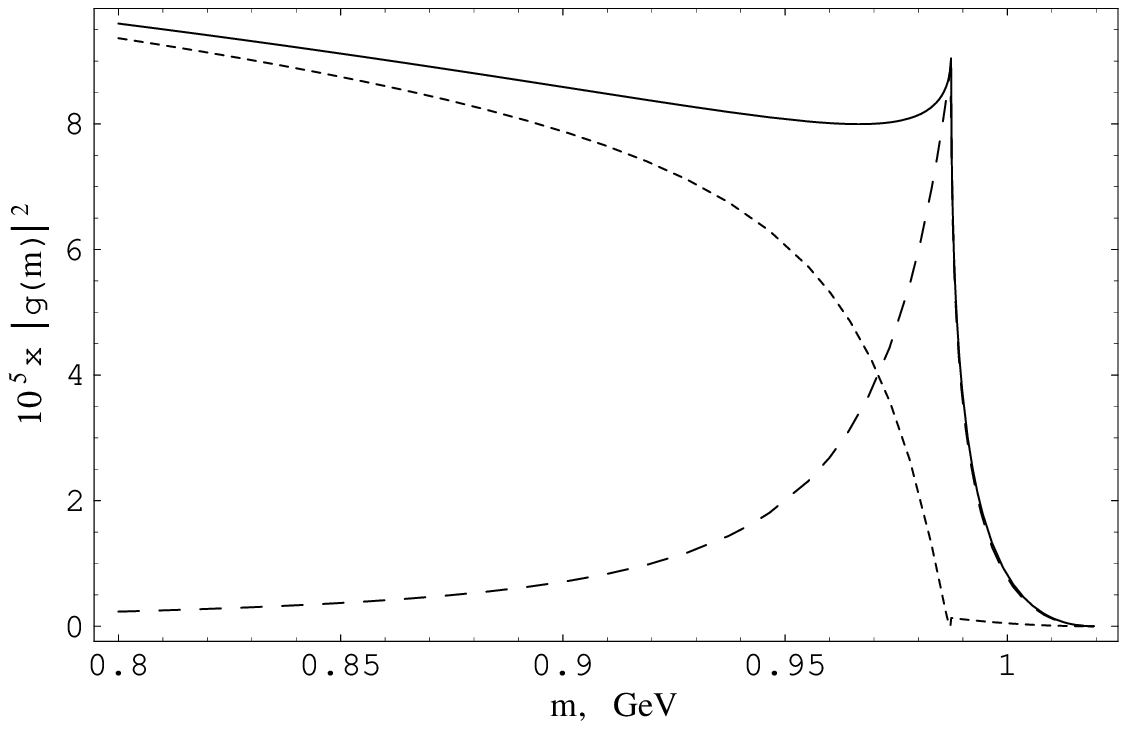}\\ (a)&(b)&(c)
\end{tabular}
\end{center}
{\footnotesize Figure 7. (a) The fit to the KLOE data for the
$\pi^0\eta$ mss spectrum in the $\phi\to\gamma\pi^0\eta$ decay.
(b) The fit to the KLOE data for the $\pi^0\pi^0$ mss spectrum in
the $\phi\to\gamma\pi^0\pi^0$ decay. (c) The universal in the
$K^+K^-$ loop model function $|g(m)|^2=|g_R(m)/g_{RK^+K^-}|^2$ is
shown by the solid curve. The contribution of the imaginary (real)
part is shown by dashed (dotted) curve.}
\end{figure}

For $dBR[\phi\to\gamma (a_0/f_0)\to\gamma
(\pi^0\eta/\pi^0\pi^0),\,m]/dm \sim |g(m)|^2\omega (m)$, the
function $|g(m)|^2$ should be smooth at $m\leq 0.99$ GeV. But
gauge invariance  requires that $g(m)$ is proportional to the
photon energy $\omega(m)$. Stopping the function $(\omega (m))^3$
at $\omega (990\,\mbox{MeV})$=29 MeV is the crucial point. The
$K^+K^-$ loop model
 solves this problem in the elegant way, see Fig. 7 (c).

\section{Four-quark transition and OZI rule \cite{A2}}
So, we are dealing here with the four-quark transition.
 A radiative four-quark transition between two $q\bar q$
states requires creation and annihilation of an additional $q\bar
q$ pair, i.e., is forbidden according to OZI rule, while a
radiative four-quark transition between $q\bar q$ and $q^2\bar
q^2$ states requires only creation of an additional $q\bar q$
pair, i.e.,  is allowed according to the OZI rule.The large $N_C$
expansion support this conclusion.
\section{Why  \boldmath $a_0(980)$ and  $f_0(980)$ are
not the $K\bar K$ molecules \cite{AK0708}}
 Every diagram  contribution in\\[9pt]
$ T\left \{\phi(p)\to\gamma [a_0(q)/f_0(q)]\right \}=
 (a)+(b)+(c)$ (see Fig. 6)\\[9pt]
is divergent and hence should be regularized in a gauge invariant
manner, for example, in the Pauli-Villars one.\\[9pt]
$\overline{T}\left \{\phi(p)\to\gamma [a_0(q)/f_0(q)], M\right\} =
\overline{(a)}+\overline{(b)}+\overline{(c)}\,,\\[9pt]
\overline{T}\left\{ \phi(p)\to\gamma [a_0(q)/f_0(q)], M
\right\}=\epsilon^\nu(\phi)\epsilon^\mu(\gamma)\overline{T}_{\nu\mu}(p,q)
=\epsilon^\nu(\phi)\epsilon^\mu(\gamma)\left[\overline{a}_{\nu\mu}(p,q)+\overline{b}_{\nu\mu}(p,q)+
\overline{c}_{\nu\mu}(p,q)\right],$\\[12pt]
$\overline{a}_{\nu\mu}(p,q)= -\frac{i}{\pi^2}\int\left
\{\frac{(p-2r)_\nu(p+q-2r)_\mu}{(m_K^2-r^2)[m_K^2-(p-r)^2][m_K^2-(q-r)^2]}
-\frac{(p-2r)_\nu(p+q-2r)_\mu}{(M^2-r^2)[M^2-(p-r)^2][M^2-(q-r)^2]}\right
\}dr\,,$\\[12pt]
 $\overline{b}_{\nu\mu}(p,q)=
-\frac{i}{\pi^2}\int\left
\{\frac{(p-2r)_\nu(p-q-2r)_\mu}{(m_K^2-r^2)[m_K^2-(p-r)^2][m_K^2-(p-q-r)^2]}-\frac{(p-2r)_\nu(p-q-2r)_\mu}{(M^2-r^2)[M^2-(p-r)^2][M^2-(p-q-r)^2]}\right
\}dr\,,$\\[12pt] $\overline{c}_{\nu\mu}(p,q)=-\frac{i}{\pi^2}\,
2g_{\nu\mu}\int dr \left \{\frac{1}{(m_K^2-r^2)[m_K^2-(q-r)^2]}-
\frac{1}{(M^2-r^2)[M^2-(q-r)^2]}\right \}\,,$\\[9pt]
 where $M$ is
the regulator field mass. $M\to\infty$ in the end\\[9pt]
$\overline{T}\bigl [\phi\to\gamma (a_0/f_0), M\to\infty\bigr ]\to
T^{Phys}\bigl [\phi\to\gamma (a_0/f_0) \bigr ]\,.$ \\[9pt]
 We can shift  the
integration variables in the regularized amplitudes and easily
check the gauge invariance condition\\[9pt]
$\epsilon^\nu(\phi)k^\mu\overline{T}_{\nu\mu}(p,q)=\epsilon^\nu(\phi)(p-q)^\mu\overline{T}_{\nu\mu}(p,q)=0\,.$\\[9pt]
 It is instructive to consider how the gauge
invariance condition\\[9pt]
$\epsilon^\nu(\phi)\epsilon^\mu(\gamma)\overline{T}_{\nu\mu}(p,p)=0$\\[9pt]
holds true,\\[9pt]
$\epsilon^\nu(\phi)\epsilon^\mu(\gamma)\overline{T}_{\nu\mu}(p,p)=
\epsilon^\nu(\phi)\epsilon^\mu(\gamma)T^{\,m_{K}}_{\nu\mu}(p,p)
-\, \epsilon^\nu(\phi)\epsilon^\mu(\gamma)T^M_{\nu\mu}(p,p)=
(\epsilon(\phi)\epsilon(\gamma))(1-1)=0\,.$\\[9pt]
 The
superscript $m_K$  refers to the nonregularized amplitude and the
superscript $M$
 refers to the regulator field amplitude.
 So, the contribution of the (a), (b), and (c) diagrams does not
depend on a  particle mass in the loops ($m_K$ or $M$) at $p=q$
\footnote{A typical example of such integrals is
$2\int_0^\infty\frac{m^2x}{(x+m^2)^3}dx=1$.}. But, the physical
meaning of  these contributions is radically different. The
$\epsilon^\nu(\phi)\epsilon^\mu(\gamma)T^{\,m_{K}}_{\nu\mu}(p,p)$
contribution is caused by  intermediate momenta (a few GeV)  in
the loops , whereas the regulator field contribution is caused
fully by high momenta ($M\to\infty $) and teaches us how to allow
for high $K$ virtualities in a gauge invariant way.

Needless to say the integrand of
$\epsilon^\nu(\phi)\epsilon^\mu(\gamma)\overline{T}_{\nu\mu}(p,p)$
is not equal to 0.

 It is clear that\\[9pt]
$\epsilon^\nu(\phi)\epsilon^\mu(\gamma)T^{M\to\infty}_{\nu\mu}(p,q)\to
\epsilon^\nu(\phi)\epsilon^\mu(\gamma)T^{M\to\infty}_{\nu\mu}(p,p)
\equiv\epsilon^\nu(\phi)\epsilon^\mu(\gamma)T^{M}_{\nu\mu}(p,p)\equiv(\epsilon(\phi)\epsilon(\gamma))\,.$\\[9pt]
  So, the  regulator field contribution
tends to the subtraction constant when $M\to\infty $.

The finiteness of the subtraction constant hides its high momentum
origin  and sometimes gives rise to an illusion of a
nonrelativistic physics in the $K^+K^-$ model with the pointlike
interaction.

\section{\boldmath $a_0(980)/f_0(980)\to\gamma\gamma$  \& $q^2\bar
q^2$ model } Twenty six years ago we  predicted \cite{ADS} the
suppression of $a_0(980)/f_0(980)\to\gamma\gamma$ decays   basing
on $q^2\bar q^2$ model. Experiment supported this prediction.  The
$a_0\to K^+K^-\to\gamma\gamma$ model \cite{AS88} describes
adequately data and corresponds to the four-quark transition
$a_0\to q^2\bar q^2\to\gamma\gamma$.\\[9pt]
 $\langle\Gamma(a_0\to
K^+K^-\to\gamma\gamma\rangle\approx 0.3$ keV.
$\Gamma^{direct}_{a_0\to\gamma\gamma} \ll 0.1$.
\section{\boldmath $\gamma\gamma\to\pi\pi$
from Belle \cite{AS05,AS08}}
 Recently, we analyzed the new high statistics
Belle data on the reactions $\gamma\gamma$\,$\to$\,$\pi\pi$  and
clarified the current situation around the $\sigma(600)$,
$f_0(980)$, and $f_2(1270)$ resonances in $\gamma\gamma$
collisions, see Fig. 9.\\[9pt]
 $\langle\Gamma(\sigma\to
\pi^+\pi^-\to\gamma\gamma)\rangle\,\approx$\,0.45\,keV,\
$\langle\Gamma (f_0\to
K^+K^-\to\gamma\gamma)\rangle\,\approx$\,0.2\,keV,\\[9pt]
$\Gamma^{direct}_{\sigma\to\gamma\gamma}\ll 0.1$\,keV,\
$\Gamma^{direct}_{f_0\to\gamma\gamma} \ll 0.1$\,keV.\\[9pt]

\begin{figure}[h]
\begin{center}
\begin{tabular}{cc}
\includegraphics[width=18pc,height=16pc]{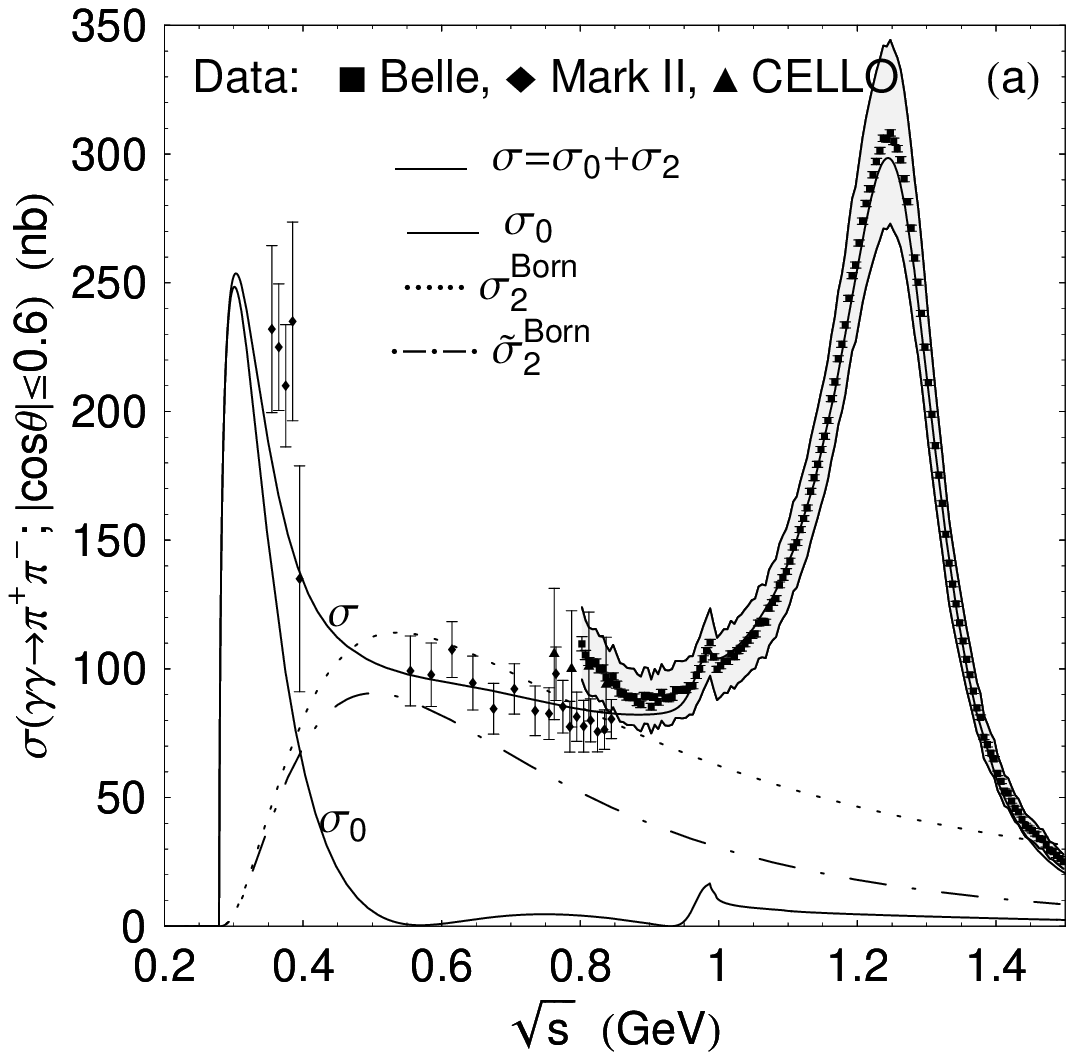}\ \
\includegraphics[width=18pc,height=16pc]{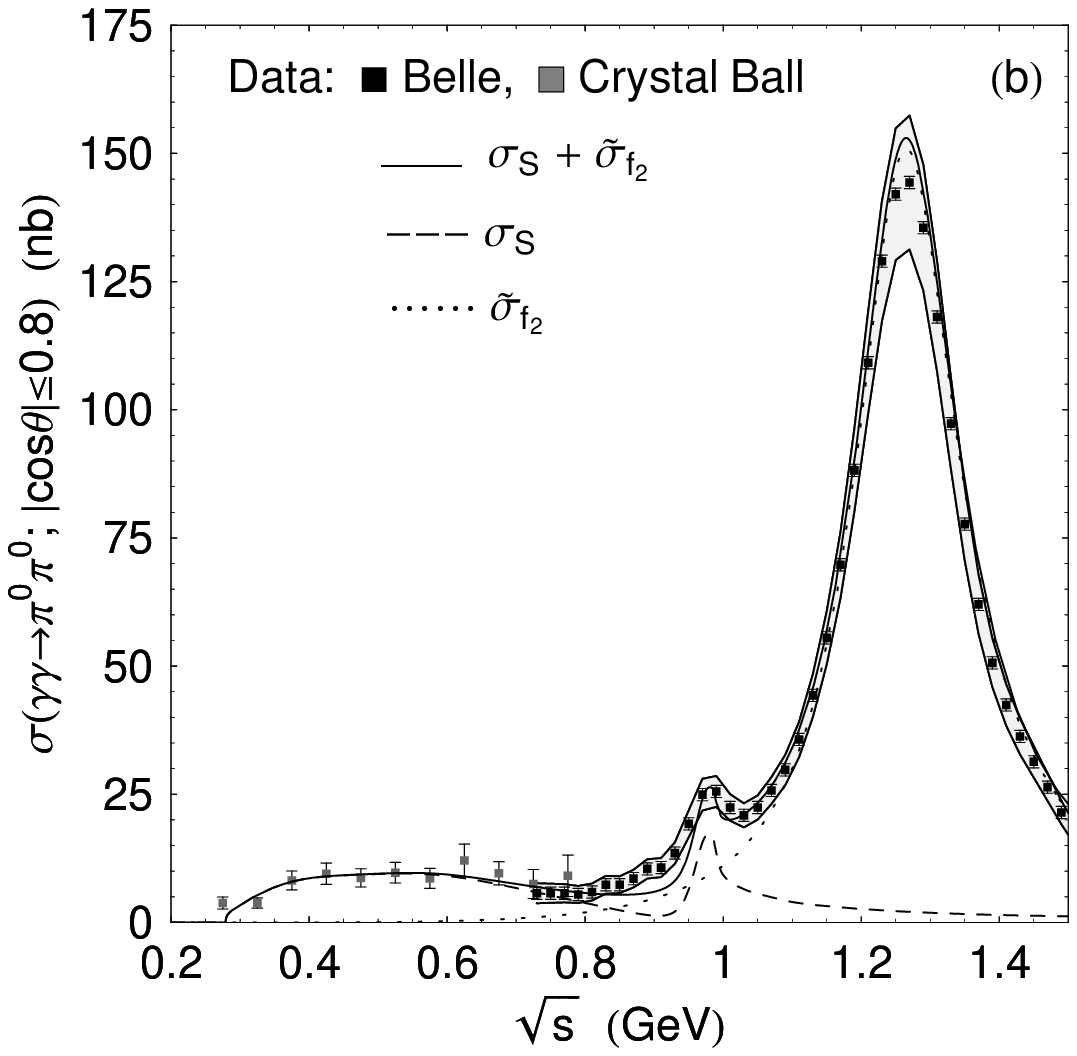}
\end{tabular}
\\ \end{center} {\footnotesize Figure 8. (a) Cross section for $\gamma\gamma\to\pi^+
\pi^-$. (b) Cross section for $\gamma\gamma\to\pi^0\pi^0$.}
\end{figure}

\section{The lessons}

The majority of current investigations of the mass spectra in
scalar channels do not study particle production mechanisms. That
is why  such  investigations are nothing more than  preprocessing
experiments, and the derivable information is very relative. And
only the progress in understanding the particle production
mechanisms can essentially advance us in revealing the light
scalar meson nature, as it is evident from the foregoing.\\[9pt]
Theoretical investigations of light scalar mesons, using effective
lagrangians  in tree level approximation, are very preliminary
ones if not exercises. Real investigations require describing real
processes $\pi\pi\to\pi\pi$, $\gamma\gamma\to\pi\pi$,
$\phi\to\gamma\pi\pi$, $\phi\to\gamma\pi^0\eta$, and so on, as it
is evident from the foregoing.\\[9pt] The $K^+K^-$ loop model,
$\phi\to K^+K^-\to\gamma (f_0/a_0)\to\gamma(\pi\pi/\pi^0\eta)$,
 describes the relativistic
physics and  strongly supports a compact $q^2\bar q^2$ nature of
$f_0(980)$ and $a_0(980)$. Consideration of Section 17 is
transferred to the  $(f_0/a_0)\to K^+K^-\to\gamma\gamma$ decays
easily. So, there are no grounds to speak about a nonrelativistic
molecular nature of $f_0(980)$ and $a_0(980)$.
\\[9pt]
 The classic $P$ wave $q\bar q$ tensor mesons
 $f_2(1270)$, $a_2(1320)$, and $f^\prime_2(1525)$ are produced by the
direct transitions $\gamma\gamma\to q\bar q$ in the main, whereas
the light scalar mesons $\sigma(600)$, $f_0(980)$, and $a_0(980)$
are produced by the rescattering
$\gamma\gamma\to\pi^+\pi^-\to\sigma$, $\gamma\gamma\to K^+K^-\to
f_0$, and $\gamma\gamma\to K^+K^-\to a_0$. The direct transitions
$\gamma\gamma\to\sigma$, $\gamma\gamma\to f_0$, and
$\gamma\gamma\to a_0$ are negligible, as it is expected in
four-quark model.

\begin{center}
 {\bf\Large Acknowledgments}
\end{center}

 I thank A.V. Kiselev,  H. Leutwyler, and G.N. Shestakov  very much
 for numerous
communications.  This work was supported in part by RFFI Grant No.
07-02-00093  and by Presidential Grant No. NSh-1027.2008.2.

\end{document}